\documentclass[twocol]{ametsocV6.1}

\usepackage{amsmath,amsfonts,amssymb,bm}
\usepackage{mathptmx}
\usepackage{newtxtext}
\usepackage{newtxmath}

\usepackage{adjustbox}
\usepackage{graphicx}

\usepackage{float}
\usepackage{booktabs}
\usepackage[]{verbatim} 	

\title{Future changes in the vertical structure of severe convective storm environments over the U.S. central Great Plains \\
{\color{red}NOT PUBLISHED. Revised version still in peer review}}
\authors{Isaac Davis\aff{a},  Funing Li\aff{b}, Daniel R. Chavas\aff{b}\correspondingauthor{Daniel R Chavas, 
    Department of Earth, Atmospheric, and Planetary Sciences, Purdue University, West Lafayette, IN. Email: dchavas@purdue.edu. }}

\affiliation{\aff{a}National Center for Atmospheric Research, Boulder, CO. \\ \aff{b} Department of Earth, Atmospheric, and Planetary Sciences, Purdue University, West Lafayette, IN }

\abstract{The effect of warming on severe convective storm potential is commonly explained in terms of changes in vertically-integrated (``bulk") environmental parameters, such as CAPE and 0--6 km shear. However, such events are known to depend on details of the vertical structure of the thermodynamic and kinematic environment that can change independently of these bulk parameters. This work examines how warming may affect the complete vertical structure of these environments for fixed ranges of values of high CAPE and bulk shear, using data over the central Great Plains from two high-performing climate models. Temperature profiles warm relatively uniformly with height, with a slight decrease in free tropospheric lapse rate, and the tropopause shifts upwards at constant temperature. The boundary layer becomes slightly drier (-2--4\% relative humidity) while the free troposphere becomes slightly moister (+2--3\%). Moist static energy (MSE) increases relatively uniformly with height with slightly larger increase within the boundary layer. Moist static energy deficit increases slightly above 4 km altitude. Wind shear and storm-relative helicity increase within the lowest 1.5 km associated with stronger hodograph curvature. Changes are broadly consistent between the two models despite differing biases relative to ERA5. The increased low-level shear and SRH suggests an increased potential for severe thunderstorms and tornadoes, while the slight increase in free tropospheric MSE deficit (enhanced entrainment) and decrease in boundary layer relative humidity (higher LCL) may oppose these effects. Evaluation of the net response of severe convective storm outcomes cannot be ascertained here but could be explored in simulation experiments.}

\begin{document}

\maketitle

%
%
%
\statement
Severe thunderstorms and tornadoes cause substantial damage and loss of life each year, which raises concerns about how they may change as the world warms. We typically use a small number of common atmospheric parameters to understand how these localized events may change with climate change. However, climate change may alter the weather patterns that produce these events in ways not captured by these parameters. This work examines how climate change may alter the complete vertical structure of temperature, moisture, and wind and discusses the potential implications of these changes for future severe thunderstorms and tornadoes.

%
%


\section{Introduction}\label{sec:introduction}




Severe convective storms (SCS) produce large hail, strong convective wind gusts, and tornadoes, all of which pose a significant threat to life and property annually \citep{Ashley_2007,Strader_Ashley_2015}. It is thus of great interest to understand how SCS activity and their associated risks may change in the future as the climate warms. Future changes in risk are difficult to assess because severe thunderstorms are too small in scale to be resolved in modern climate models. Nonetheless, recent research using downscaling simulations that can resolve thunderstorm systems have found that proxies for SCS activity show broad increases in frequency and severity in the future over the United States as the climate warms \citep{Ashley_Haberlie_Gensini_2023,Trapp_Hoogewind_LasherTrapp_2019,Hoogewind_Baldwin_Trapp_2017,Gensini_Mote_2015}, with notable shifts in their spatial distribution and seasonal cycle.

Understanding these future changes in SCS activity is rooted in our understanding of how SCS outcomes are linked to the larger-scale thermodynamic environment within which these storms are generated \citep{Ashley_Haberlie_Gensini_2023}. Environments favorable for SCS activity (``SCS environments") are typically defined by the combination of a thermodynamic ingredient, given by convective available potential energy (CAPE), and a kinematic environmental ingredient, given by 0--6 km bulk wind difference (S06; ``bulk shear''). Tornado-favorable environments are defined using a third environmental parameter, storm-relative helicity, that is often calculated within the lowest 1 km \citep{Coffer_etal_2019,Coffer_Taszarek_Parker_2020}. These parameters have been widely used to explain the historical spatiotemporal pattern of severe thunderstorms and tornadoes, particularly over the United States \citep{Gensini_Brooks_2018,Taszarek_etal_2021,Coffer_Taszarek_Parker_2020,Li_etal_2021,Hoogewind_Baldwin_Trapp_2017,chen2020} but also globally \citep{Brooks_Lee_Craven_2003,Allen_Karoly_Mills_2011,Taszarek_etal_2020,Taszarek_etal_2021}. Note that these parameters represent necessary but not sufficient conditions for severe convective storms and tornadoes, as a triggering mechanism for convective initiation is also required, though this step is relatively poorly understood and hence not easily incorporated into environmental analyses \citep[e.g.,][]{Ashley_Haberlie_Gensini_2023}. In future climate projections, the increase in SCS activity over North America is explained principally by large projected increases in CAPE \citep{Ashley_Haberlie_Gensini_2023,Tippett_etal_2015, Gensini_2021, Lepore_etal_2021, Seeley_Romps_2015, Diffenbaugh_etal_2013, Hoogewind_Baldwin_Trapp_2017}, consistent with a theoretical model that predicts a rapid increase in continental diurnal-maximum CAPE with warming \citep{Agard_Emanuel_2017}. Finally, tropospheric relative humidity is expected to remain relatively constant with warming based on observations and climate models \citep{Douville_etal_2022}, though this outcome has yet to be evaluated specifically in the context of SCS environments.


However, each of these environmental parameters are ``bulk'', i.e., vertically-integrated measures of buoyancy or shear. Substantial research has shown there are pathways to change SCS outcomes that are independent of these bulk measures. This includes dependencies on the vertical distribution of buoyancy and shear \citep{McCaul_Weisman_2001}, low-level shear profile \citep{Guarriello_Nowotarski_Epifanio_2018,Peters_etal_2023a},
low-level thermal and moisture structure \citep{Mccaul_Cohen_2002,Brown_Nowotarski_2019}, and hodograph curvature \citep{Nixon_Allen_2022}. Additionally, SCS outcome is sensitive to free-tropospheric relative humidity \citep{Chavas_Dawson_2021,Lashertrapp_etal_2021,Jo_Lashertrapp_2022}, due in part to its strong effects on entrainment dilution that reduces parcel buoyancy and CAPE \citep{Peters_etal_2020b}; however, CAPE itself is nearly insensitive to this quantity.

Thermodynamic and kinematic profiles are complex, which makes it difficult to define and interpret variations in their vertical structure. Recently, \cite{Chavas_Dawson_2021} developed a simple model for the SCS environmental thermodynamic and kinematic sounding structure comprised of a boundary layer model and a free-tropospheric model. The model is consistent with how SCS environments are generated downstream of the Rocky Mountains \citep{Carlson_Ludlam_1968,Doswell_2001,Agard_Emanuel_2017}. This sounding framework offers a foundation for defining the vertical structure of an SCS environment based on a relatively small number of parameters. Hence, the model may also provide a useful framework for defining key changes in vertical structure in the future.

How the complete vertical structure of SCS environments, rather than simply bulk parameters, may change in a future climate has received relatively little attention to date.
To fill this knowledge gap, we seek to answer the following research questions:
\begin{enumerate}
\item How do the vertical thermodynamic and kinematic structure of severe convective storm environments change in high-performing CMIP6 climate model projections over the central Great Plains?
\item Are the changes consistent between two high-performing climate models?
\end{enumerate}
To answer these questions, we investigate how the vertical thermodynamic profiles (temperature, relative humidity and moist static energy deficit) and kinematic profiles (wind shear and storm relative helicity) in severe storm environments may change with future warming. We focus here on the central Great Plains, which is within the primary severe thunderstorm and tornado hotspot over North America, to minimize geographic and seasonal variability. There is evidence of an eastward shift towards the southeast U.S. in recent decades \citep{Gensini_Brooks_2018}, though the nature and dynamics of SCS environments are known to differ in this region. As a result, mixing the two regions is not advisable, but future work may seek to expand this analysis to the southeast U.S. For our analysis, we take advantage of recent work that identified a small number of climate models that best reproduce the historical SCS environment climatology over North America \citep{Chavas_Li_2022}. We further use a simple model for the SCS environmental sounding from \cite{Chavas_Dawson_2021} as a guiding framework to define key parameters that capture the basic vertical thermodynamic and kinematic structure, though our results are not specific to that model in order to keep our findings general. 

We detail our methodology in Section 2. We present our results for future changes in the structure of SCS environments and discuss key outcomes in section 3. Finally, we summarize conclusions and avenues for future work in section 4.

\section{Methodology}\label{sec:methdology}

\begin{figure*}[!ht]
\centerline{\includegraphics[width=0.9\textwidth]{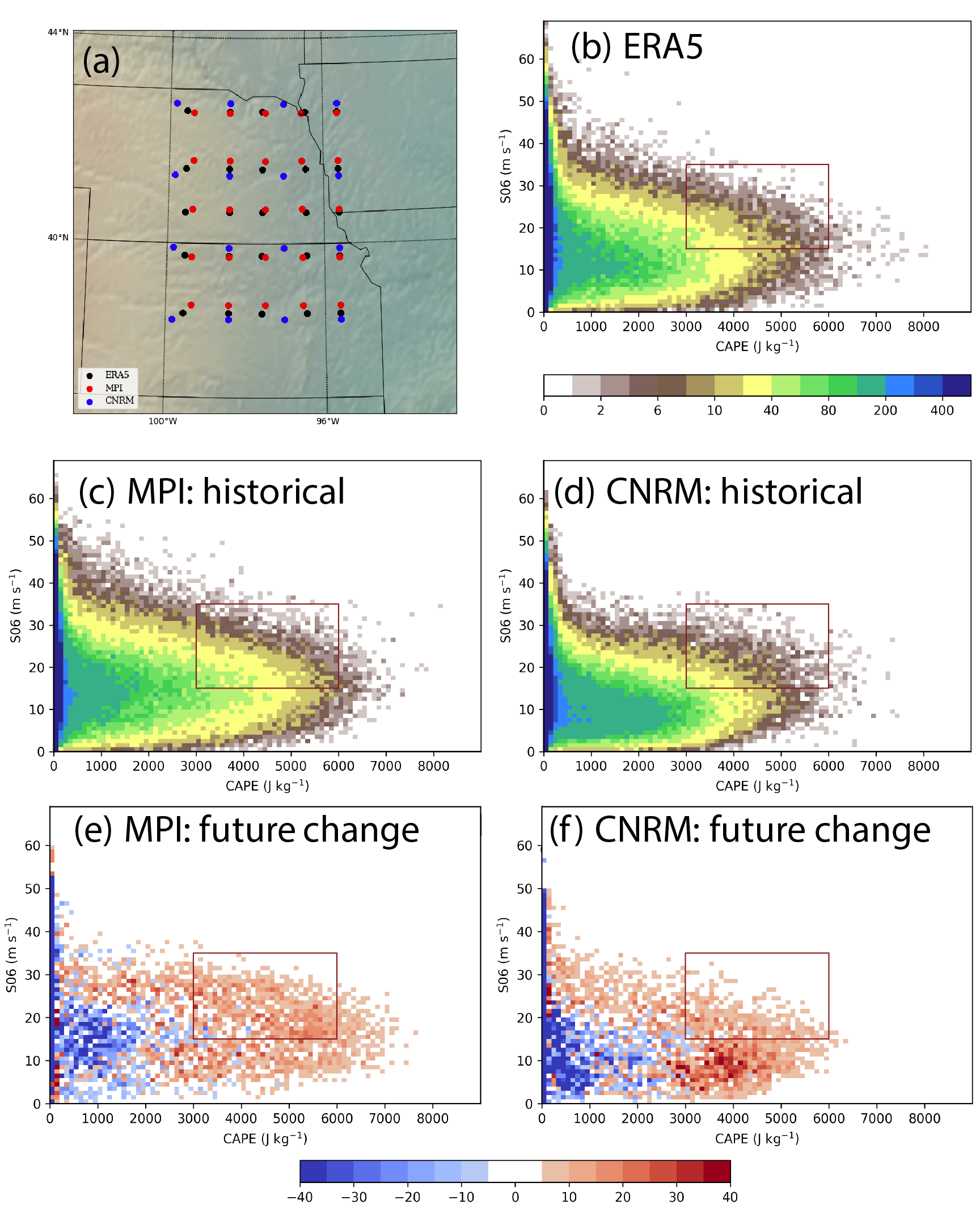}}    
\caption{(a) Map of gridpoint distributions within our region of interest from the ERA5 historical dataset and from the MPI and CNRM climate model datasets. (b) Joint histogram of CAPE and bulk shear (S06) from the ERA5 dataset for March-June for the period 1980--2014, with box denoting the fixed ranges of CAPE and S06 examined in this study. (c)-(d) As in (b) but for MPI and CNRM, respectively. (e)-(f) Same as (c)-(d) but for the difference between the future ssp370 (2065--2099) and historical periods for each model. CNRM values have been multiplied by the factor 25/16 to account for its smaller number of gridpoints for an apples-to-apples comparison with ERA5 and MPI.}
\label{fig:models}
\end{figure*}

This work uses reanalysis and climate model data to examine future thermodynamic and kinematic changes in vertical structure independent of changes in bulk SCS environmental parameters (CAPE and S06). To do so, we focus our analysis on a limited geographic region, to minimize regional variability in the climatology \citep{Taszarek_etal_2020b}, and within a fixed range of values of CAPE and S06, to minimize effects of future changes in CAPE and S06 itself. We use soundings from March through June in a region over eastern Kansas and Nebraska within the central Great Plains bounded by 100--95 $^o$W and 38--43 $^o$N (Figure \ref{fig:models}a), which is the same subdomain used in \cite{Li_etal_2020}, to focus on the primary severe convective storm season. We retain soundings with CAPE between 3000 and 6000 J kg$^{-1}$ and S06 between 15 and 35 ms$^{-1}$ to capture the principal regime of CAPE-S06 values associated with significant SCS activity in the historical record \citep{Brooks_Lee_Craven_2003,Li_etal_2020}. We further impose an upper bound on the magnitude of convective inhibition of CIN$<$125 J kg$^{-1}$ following \citep{Lepore_etal_2021} to avoid environments unlikely to allow convective initiation, though results are similar without this criteria; its inclusion eliminates relatively few soundings as shown below. For both the thermodynamic and kinematic vertical structure, we first compare ERA5 data to climate model historical experiments to examine similarities and biases of the models. We then analyze changes between the future and historical period for each climate model. 

For climate model analysis, we select two specific CMIP6 climate models \citep{Eyring_etal_2016}: MPI-ESM1-2-HR (hereafter ``MPI'') and CNRM-ESM2-1 (hereafter ``CNRM''). These two models were identified by \cite{Chavas_Li_2022} as the two highest performing models in the CMIP6 archive for reproducing the historical SCS environment climatology over North America. \cite{Chavas_Li_2022} demonstrated that climate models exhibit a very wide range of variability in the climatological representation of severe convective storm environments (spatiotemporal pattern and amplitude) relative to historical data, and thus it is important to first select models that can credibly reproduce the historical record. Though both models perform comparably well in reproducing the overall SCS environment climatology over North America, they still may differ in their representation of such environments specifically over our central Great Plains region of interest, as will be noted below. This outcome can be useful to check whether the models yield consistent responses in vertical structure despite differing mean-state biases, which suggests greater robustness; this is a common approach when working with climate models for which the mean state (e.g. global-mean temperature) can vary across models but the structure of their responses to forcing (atmospheric ``fingerprint'', e.g. spatial structure of warming due to increased greenhouse gas concentrations) may be quite similar \citep{Santer_etal_2013,Zhang_etal_2023}. We use the radiative forcing ssp370 experiment for each model as our future simulation, which is considered as a more plausible high-end forcing scenario than ssp585 \citep{Pielke_Burgess_Ritchie_2022}. Results for ssp585 are qualitatively similar (not shown). The historical period used is 1980--2014 and future period is 2065--2099.

We compare climate model output against the ERA5 reanalysis model-level data for the identical period to match the climate model historical period \citep{Hersbach_etal_2020}. Model level data ensures use of the highest vertical resolution data available, though results are similar when using pressure-level data (not shown). ERA5 is sampled at 00, 06, 12, 18 UTC to match the same 6-hourly output available from both climate models. ERA5 is the highest resolution of existing global long-term reanalysis datasets and performs well in reproducing the climatological spatiotemporal distribution of SCS environments found in radiosonde observations \citep{Li_etal_2020}. The vertical resolution of ERA5 (137 levels) and the two climate models (95 levels for MPI, 91 levels for CNRM) differ, so for direct comparison between ERA5 and the climate model data we linearly interpolate the model data in the vertical to match ERA5. The horizontal resolution for ERA5 is $\Delta x =\sim$31 km, $\Delta x = \sim$100 km for MPI, and $\Delta \sim$x = 140 km for CNRM. We downsample ERA5 to every fourth gridpoint to more closely match the spacing of the models; the grids for each within our domain of interest are displayed in Figure \ref{fig:models}a. ERA5 and MPI both contain 25 grid points with the resolution of MPI being slightly finer, while CNRM contains 16 grid points. In contrast to the vertical grid, we do not interpolate data horizontally to the same grid to avoid mixing soundings at adjacent grid points that represent very different environments, such as in the vicinity of a cold front, which is not uncommon for SCS environments.

For a given sounding, we define the tropopause as the lowest altitude where the lapse rate drops below 2 K km$^{-1}$ \citep{slownik1992, Chavas_Li_2022}, and define the top of boundary layer as the height of maximum relative humidity following \cite{Chavas_Dawson_2021}. We calculate CAPE, given by
\begin{equation} CAPE={\int_{z_{LFC}}^{z_{EL}} g\frac{T_{vp}-T_{ve}}{T_{ve}} dz}, \end{equation}
where $g=9.81$ is the acceleration due to gravity; $z$ is altitude with subscripts `LFC' and `EL' denoting level of free convection and equilibrium level, respectively; and $T_v$ is the virtual temperature with subscripts `p' and `e' denoting parcel and environment, respectively. CAPE is calculated using the xcape codebase \citep{Lepore_Allen_Abernathey_2021_dataset} for the near-surface (z = 2 m) parcel as in \cite{Chavas_Li_2022}. Bulk wind shear within a layer between bottom altitude $z_b$ and top altitude $z_t$ is given by the magnitude of the vector wind difference across the layer
\begin{equation} S[z_b][z_t]=|\textbf{V}_{z_t}-\textbf{V}_{z_b}|, \end{equation}
For the standard 0-6 km shear layer (S06), $z_b$ = 10 m and $z_t$ = 6000 m. Finally, storm-relative helicity (SRH) is given by
\begin{equation} SRH_{0-z_t}={\int_{10m}^{z_t} (\textbf{V}-\textbf{C})\cdot\left(\nabla \times \textbf{V}\right) dz}, \end{equation}
where $\textbf{V}$ is the wind vector at a given level, $\textbf{C}$ is the storm-motion vector, and $z_t$ is defined in the same manner as with shear. In our results below, we analyze wind shear variations between the surface and 6 km and SRH integrated over layers between the surface and 3 km. All parameters involving wind (hodographs, wind shear, SRH) are calculated using the hodograph, storm relative helicity, and Bunkers storm motion functions from MetPy v1.1 \citep{metpy}.

The climatological joint histogram of CAPE and S06 for ERA5 is shown in Figure \ref{fig:models}b and for the historical periods in MPI and CNRM, respectively, in Figure \ref{fig:models}c--d. Our fixed ranges of high CAPE and S06 values are highlighted by the red box, which is the same in each subplot. Because CNRM contains fewer grid points than ERA5 and MPI, we rescale the histogram data for CNRM by the factor 25/16 to account for its smaller number of gridpoints for an apples-to-apples comparison across all datasets.  Both models do very well in reproducing the structure of the climatological joint distribution of CAPE and S06. Relative frequencies peak at moderate S06 (10-15 $m \: s^{-1}$) and small values of CAPE, and the largest CAPE values are associated with these moderate values of S06, consistent with past work \citep{Li_etal_2020,Taszarek_etal_2020}. Within our phase space subset of interest (the box), CNRM captures both the structure and magnitude very well, while MPI captures the structure but its magnitude has a clear high bias. Future changes in this joint distribution relative to historical within each model are shown in Figure \ref{fig:models}e--f. In the future, in both models the phase space distribution shifts principally rightwards, associated with an increase in the frequency of relatively high CAPE ($>$3000 J kg$^{-1}$) but minimal change in S06 (Figure \ref{fig:models}e--f), again consistent with past research \citep{Lepore_etal_2021}. Within our phase space subset of interest, this shift represents an increase in the absolute frequency of such environments (predominantly red colors). Overall, then, both models perform reasonably well in reproducing the climatology of SCS environments, though CNRM better captures the magnitude, consistent with the findings of \cite{Chavas_Li_2022}.

\begin{figure*}[!ht]
\centerline{\includegraphics[width=\textwidth]{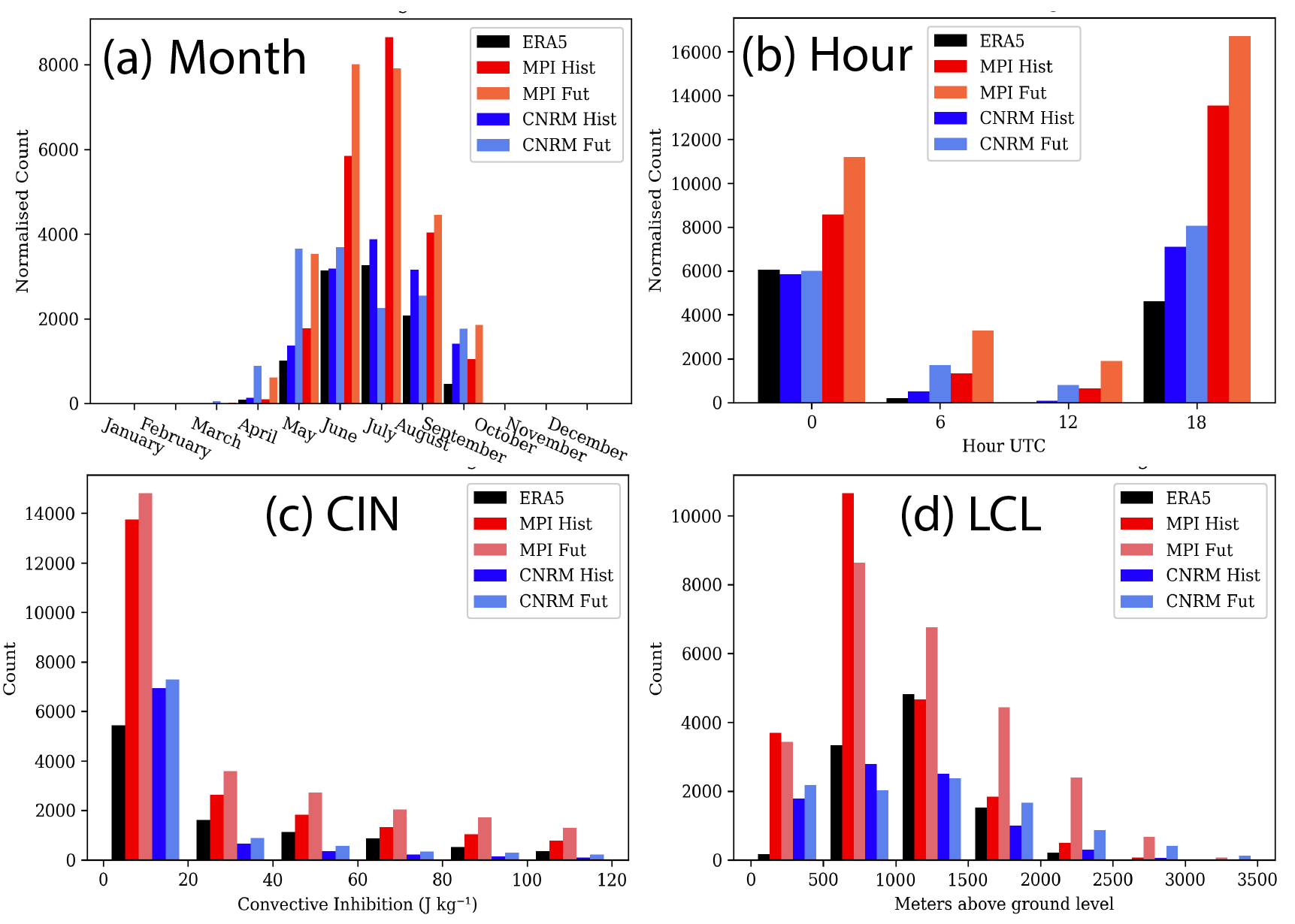}}    
\caption{(a) Monthly frequency of final subset for ERA5, MPI historical and ssp370 future, and CNRM historical and ssp3 future. (b) Diurnal frequency of final subset (00/06/12/18 UTC). (c) Frequency distribution of convective inhibition (CIN) within final subset in 20 J kg$^{-1}$ bins starting from zero. (d) Frequency distribution of lifted condensation level (LCL) within final subset in 500 m bins starting from zero. CNRM values have been multiplied by the factor 25/16 to account for its smaller number of gridpoints for an apples-to-apples comparison with ERA5 and MPI.}
\label{fig:monthlydiurnal}
\end{figure*}

The monthly and diurnal frequency distributions of our final sounding datasets are shown in Figure \ref{fig:monthlydiurnal}. In ERA5, the seasonal cycle of SCS environments (Figure \ref{fig:monthlydiurnal}a) closely follows the seasonal cycle of observed severe thunderstorm hazards and tornadoes for this region \citep[Figures 5--6 of][]{Taszarek_etal_2020b}, with a rapid increase from March through June, a peak in June/July, then a rapid decrease from July through October. CNRM does very well in reproducing the ERA5 seasonal cycle structure and amplitude, particularly for April through July, with a moderate high bias in August and September. Meanwhile, MPI also captures the structure but exhibits a strong high bias in magnitude that is consistent throughout the season. In the future, both models exhibit a shift in the seasonal cycle towards earlier months (April/May) as has been found in recent work \citep{Ashley_Haberlie_Gensini_2023}. 

The diurnal cycle (Figure \ref{fig:monthlydiurnal}b) is skewed strongly towards afternoon/evening as expected, with the vast majority of soundings in ERA5 at 00 UTC ($\sim 55\%$) and 18 UTC ($\sim 45\%$) and very few events at 06 UTC and 12 UTC. While 18 UTC may seem early relative to the typical timing of convective storm occurrence in the late afternoon \citep[Figure 8 of][]{Taszarek_etal_2020b}, this difference likely reflects a lag between the gradual daytime generation of these soundings and the initiation of convection itself. CNRM captures the ERA5 diurnal cycle structure and amplitude well, though its distribution is shifted towards 18 UTC, i.e., a bias towards too early in the day. MPI again captures the gross diurnal cycle structure with an overall strong high bias in magnitude, but notably it exhibits a similar timing bias towards 18 UTC as CNRM. This early timing bias may be associated with the known and early timing bias in inland precipitation itself that has been persistent in climate models \citep{Christopoulos_Schneider_2021}. In the future, the distribution of soundings across the diurnal cycle remains relatively constant.

Finally, we examine the frequency distributions of convective inhibition (CIN; Figure \ref{fig:monthlydiurnal}c) and of lifted condensation level (LCL; Figure \ref{fig:monthlydiurnal}d). CIN frequency peaks at relatively small values in ERA5 as well as in both climate models. There is a long tail of relatively low frequencies for higher CIN values up to our upper-bound threshold, and as a result this criterion is already met for nearly all soundings that met our CAPE and S06 thresholds. MPI yields similar relative frequencies to ERA5, while CNRM is skewed more strongly towards smaller values. Both models reproduce the CIN distributions within SCS environments found in ERA5 relatively well. In the future, the relative frequency shifts towards slightly higher CIN, as the frequency of the lowest CIN bin increases by a significantly smaller percentage ($+10\%$) than the higher bins (e.g. $+25\%$ for 60-80 J kg$^{-1}$). This increase in CIN with warming has been found in numerous past downscaling studies \citep[e.g.][]{Ashley_Haberlie_Gensini_2023}. 

As for the LCL, in ERA5 the frequency distribution peaks at 1000-1500 m though with relatively high frequency within 500-1000 m as well. Both models are biased towards slightly lower LCLs, with slightly higher frequencies at 500-1000 m in CNRM and a more pronounced low-LCL bias in MPI. Both climate models show a shift in their relative frequencies towards higher LCLs, suggesting a shift towards drier boundary layer as will be found below.

Overall, the above examination of the spatial, seasonal, and diurnal variability lends credence that our sounding database is broadly representative of environments favorable for severe convective storms, and that our two climate models perform reasonably well in reproducing these environments. For our region, CNRM appears to be the slightly better model given its much smaller amplitude bias. As noted above, the contrasting mean-state amplitude biases provide useful context for our analysis, as responses of the vertical structure to forcing that are consistent between the two models suggest they are less likely to be sensitive to biases in the climate model mean state.





\section{Results}\label{sec:results}



\subsection{Changes in thermodynamic vertical structure}

\subsubsection{Temperature}

\begin{figure*}[!ht]
\centerline{\includegraphics[width=0.8\textwidth]{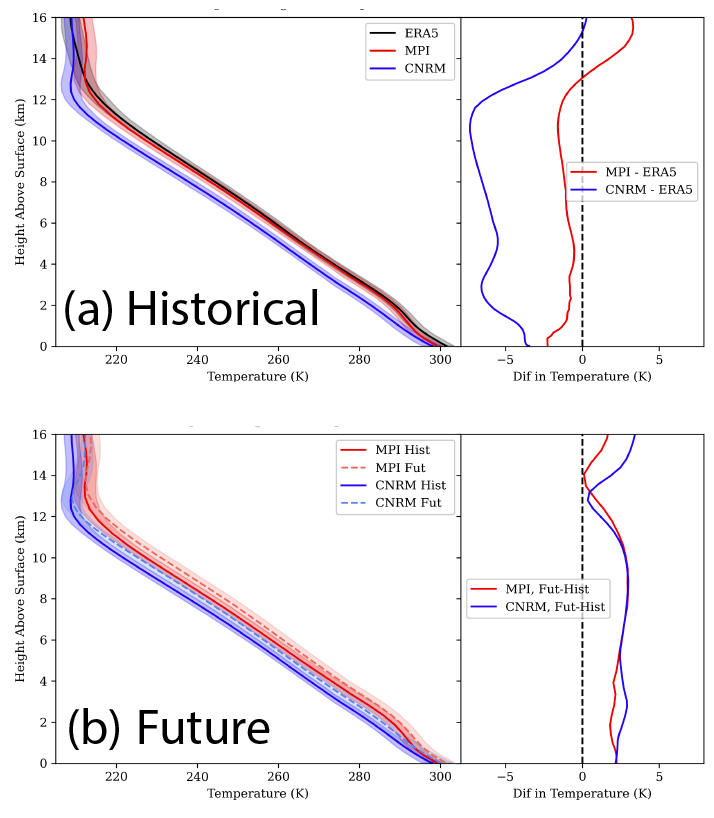}}
\caption{(a) Mean vertical profiles of temperature from ERA5 and the historic runs of MPI and CNRM. The shaded area represents the interquartile range (25th--75th percentile), and the difference plot shows the differences between the ERA5 and MPI, and ERA5 and CNRM. (b) As in (a) but for the historic and future runs of MPI and CNRM. The difference plot shows the differences between the future and historic simulations of each model.}
\label{fig:T_profile}
\end{figure*}

We begin by examining changes in thermodynamic vertical structure simulated by both models. For all of the analyses below, values of some key quantities, including those relevant to the framework of \cite{Chavas_Dawson_2021}, are provided in Table \ref{table:thermo_table} for additional reference.

Mean temperature profiles from both models for the historical climate are compared with ERA5 in Figure \ref{fig:T_profile}a. MPI performs well in reproducing the vertical structure of temperature found in ERA5, with temperatures decreasing rapidly with height above the surface within a well-mixed boundary layer, then decreasing more slowly above the boundary layer before decreasing more rapidly again through the free troposphere. The model has a slight cool bias of 1--2 K that is relatively constant with height through the depth of the troposphere. CNRM also reproduces the vertical structure of the temperature profile, though with a stronger overall cool bias of 6 K that is relatively constant with height within the free troposphere; the bias is smaller (3.5 K) within the boundary layer. The result that MPI better captures mean state temperature is consistent with \cite{Chavas_Li_2022}. While these contrasts in gross temperature bias between models may seem surprising, it is common for climate models to have differing mean state biases yet exhibit similar structural responses to forcing, as noted earlier. 

Despite different magnitudes of mean bias, both models capture the mean lapse rate within the free troposphere found in ERA5. MPI and CNRM both give lapse rate of 7.4 K km$^{-1}$ for 2--6 km and 7.6 K km$^{-1}$ for 2--10 km (Table \ref{table:thermo_table}), which closely match the values found in ERA5. Moreover, both models reasonably capture the tropopause temperature, with values of 212.2 K and 209.2 K for MPI and CNRM, respectively, compared with the ERA5 value of 212.0 K. Both models also capture the magnitude of variability in temperature, with a relatively small variability of less than 3 K magnitude in the interquartile range throughout the troposphere (Figure \ref{fig:T_profile}a). The primary qualitative difference in vertical thermal structure between the models is that MPI better captures the transition layer of reduced lapse rate separating the boundary layer and free troposphere found in ERA5.

Future changes in the simulated temperature profiles for each model are shown in Figure \ref{fig:T_profile}b. Both models project future thermal structures that are qualitatively similar to their historic simulations but simply shifted nearly uniformly warmer through the depth of the troposphere by approximately 2K. CNRM warms slightly more than MPI between 1 km and 5 km, while both models warm nearly identically in the upper free troposphere. Free tropospheric lapse rates decrease very slightly (Table \ref{table:thermo_table}), by 0.2 K km$^{-1}$ in MPI and by 0.1 K km$^{-1}$ in CNRM. In both models, the tropopause temperature remains nearly constant, while the tropopause height shifts upwards. This upward expansion at fixed tropopause temperature is consistent with a similar finding for both the midlatitudes and the tropics in general \citep{Singh_OGorman_2012,SeeleyJeevanjeeRomps2019,ThompsonCeppiLi2019}.  Moreover, the thermal structure sharpens around the tropopause including a sharper temperature increase above the tropopause, which may have impacts on the depth and strength of overshooting tops \citep{ONeill_etal_2021}.

Overall, despite some significant differences in historical vertical thermal structure biases between the two models, their future climate responses are strikingly similar. This outcome gives greater confidence in the ability to use these models to quantify future changes in the thermal structure of SCS environments.

\begin{table}[]

\caption{Key quantities calculated from the mean thermodynamic profiles from ERA5, MPI, and CNRM, including those relevant to the SCS sounding model of \cite{Chavas_Dawson_2021}. Values are the mean with interquartile range (25th-75th percentile) in parentheses. Range not included for BL height as this quantity can be noisy for individual cases.}
\centering
\begin{adjustbox}{max width=0.5\textwidth}
\begin{tabular}{|ll|l|l|l|l|l|l|l|}
\hline
 &  & ERA5 &  & \multicolumn{2}{l|}{MPI-ESM1-2-HR} &  & \multicolumn{2}{l|}{CNRM-ESM2-1} \\ \cline{5-6} \cline{8-9} 
 &  &  &  & HIST & FUT &  & HIST & FUT \\ \cline{1-3} \cline{5-6} \cline{8-9} 
\multicolumn{2}{|l|}{2--6 km Lapse Rate (K km$^{-1}$)} & 7.4 (7.0, 7.9) &  & 7.4 (7.1, 7.8) & 7.2 (6.8, 7.6) &  & 7.4 (7.0, 7.8) & 7.3 (6.8, 7.8) \\ \cline{1-3} \cline{5-6} \cline{8-9} 
\multicolumn{2}{|l|}{2--10 km Lapse Rate (K km$^{-1}$)} & 7.5 (7.3, 7.7) &  & 7.6 (7.4, 7.8) & 7.4 (7.2, 7.6) &  & 7.6 (7.4, 7.8) & 7.5 (7.3, 7.8) \\ \cline{1-3} \cline{5-6} \cline{8-9} 
\multicolumn{2}{|l|}{Tropopause Temp (K)} & 212.0 (209.7, 214.2) &  & 212.2 (209.6, 214.9) &  212.8 (210.0, 215.4) &  & 209.2 (207.0, 211.3) & 209.3 (206.9, 211.6) \\ \cline{1-3} \cline{5-6} \cline{8-9} 
\multicolumn{2}{|l|}{Tropopause Height (km)} & 13.2 (13.1, 13.3) &  & 13.2 (13.6, 13.8) & 13.3 (13.7, 13.9) &  & 12.2 (12.6, 12.8) & 12.7 (13.1, 13.3) \\ \cline{1-3} \cline{5-6} \cline{8-9} 
\multicolumn{2}{|l|}{0--1 km avg RH (\%)} & 64.3 (60.4, 68.1) &  & 76.1 (74.1, 78.4) & 72.2 (69.9, 75.0) &  & 68.9 (65.6, 72.3) & 67.7 (64.3, 71.0) \\ \cline{1-3} \cline{5-6} \cline{8-9} 
\multicolumn{2}{|l|}{1--2 km RH Lapse Rate (\% km$^{-1}$)} & 25.0 (6.5, 42.3) &  & 24.9 (10.4, 38.8) & 20.2 (6.2, 33.3) &  & 8.0 (-2.4, 18.1) & 8.3 (-2.8, 19.1) \\ \cline{1-3} \cline{5-6} \cline{8-9} 
\multicolumn{2}{|l|}{2--6 km Avg RH (\%)} & 36.2 (35.3, 37.3) &  & 43.4 (41.8, 45.2) & 46.4 (44.6, 48.3) &  & 57.7 (53.9, 62.5) & 59.1 (57.6, 60.5) \\ \cline{1-3} \cline{5-6} \cline{8-9} 
\multicolumn{2}{|l|}{2--10 km Avg RH (\%)} & 34.0 (31.0, 36.2) &  & 44.5 (43.7, 45.6) & 47.3 (45.7, 48.8) &  & 53.3 (47.8, 59.5) & 55.0 (50.3, 60.3) \\ \hline
\multicolumn{2}{|l|}{BL Height (km)} & 0.8  &  & 0.4 & 0.7 &  & 0.8  & 0.9 \\ \hline

\multicolumn{2}{|l|}{BL MSE (kJ kg$^{-1}$) } & 340.4 (337.1, 344.0) &  & 341.6 (338.1, 345.3) & 345.9 (341.7, 350.4) &  & 332.3 (329.2, 335.7) & 337.6 (334.1, 341.0) \\ \hline
\multicolumn{2}{|l|}{BL DSE (kJ kg$^{-1}$)} & 308.0 (306.2, 309.9) &  & 305.6 (303.8, 307.3) & 307.8 (305.5, 310.0) &  & 304.3 (302.0, 306.5) & 306.7 (303.8, 309.3) \\ \hline

\end{tabular}
\end{adjustbox}
\label{table:thermo_table}
\end{table}

\subsubsection{Relative humidity}


\begin{figure*}[!ht]
\centerline{\includegraphics[width=0.8\textwidth]{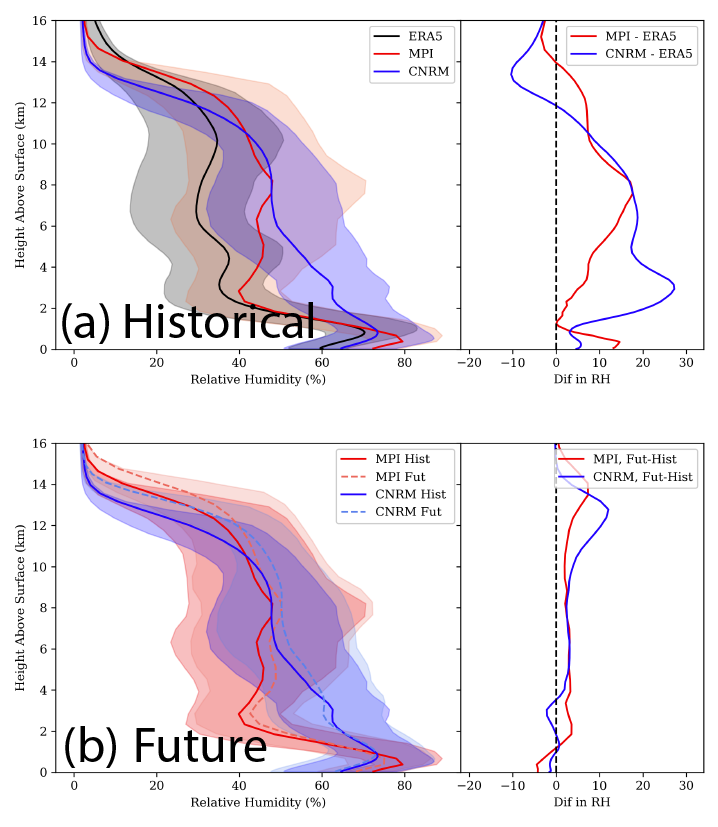}}

\caption{(a) Mean vertical profiles of relative humidity from ERA5 and the historic runs of MPI and CNRM. The shaded area represents the interquartile range (25th--75th percentile), and the difference plot shows the differences between the ERA5 and MPI, and ERA5 and CNRM. (b) As in (a) but for the historic and future ssp370 runs of MPI and CNRM. The difference plot shows the differences between the future and historic simulations of each model.}
\label{fig:RH_profile}
\end{figure*}

We next examine the vertical structure of relative humidity (RH) in Figure \ref{fig:RH_profile}. Mean RH profiles from both models for the historical climate are compared with ERA5 in Figure \ref{fig:RH_profile}a. MPI performs well in reproducing the gross vertical structure of RH found in ERA5, with RH increasing with height above the surface toward a local maximum near the top of the well-mixed boundary layer, then decreasing sharply up to approximately 2.5 km altitude before becoming relatively constant with height through the rest of the free troposphere. The model is moister than ERA5, with a moist RH bias of 12\% within the lowest 1 km, near-zero bias through the 1--2 km transition layer, and a moist RH bias of 5--15\% within the free troposphere (7\% for 2--6 km mean; 10\% for 2--10 km mean). MPI does capture the mean 1--2 km RH lapse rate of 25\% found in ERA5. CNRM also reproduces the vertical structure of the RH profile, though with a smaller boundary layer bias and larger moist bias through both the transition layer and lower free troposphere below 6 km, with a peak moist bias of 27\% at 3 km altitude. This latter moist bias indicates a much slower decrease in RH between the moist boundary layer and drier free troposphere (1--2 km RH lapse rate of 8.0\% as compared to 25\% for ERA5). This behavior is consistent with the weaker thermal transition layer (Figure \ref{fig:T_profile}a), which suggests stronger shallow convective mixing through the top of the boundary layer and into the lower free troposphere in CNRM \citep[e.g.,][]{Hu_etal_2022}.

Note that there is a much larger range of variability in RH across soundings relative to temperature (compare width of interquartile shading with Figure \ref{fig:T_profile}a). These ranges of variability are also well-captured by the models. This distinction is a clear indication of how free tropospheric moisture can vary widely on short spatial and temporal scales, as its local structure depends on the transport by antecedent convection and the mesoscale and synoptic scale flow, all of which may vary strongly in space and time. In contrast, free tropospheric temperatures are much more strongly constrained by the larger-scale dynamics of the mid-latitude atmosphere.

Despite quite different magnitudes of mean bias, both models do capture the gross vertical structure of RH, with the primary contrast in the transition layer between the boundary layer and the free troposphere. Moreover, both models reproduce the magnitude of variability in RH.

Future changes in the simulated RH profiles for each model are shown in Figure \ref{fig:RH_profile}b. Both models project future thermal structures that remain qualitatively similar to their historic simulations but shifted to slightly higher RH through the middle and upper free troposphere and slightly lower RH in the boundary layer. In the free troposphere, MPI and CNRM RH increases by 3.0\% and 1.5\% for the 2--6 km mean, respectively, and 2.8\% and 1.7\% for the 2--10 km mean, respectively (Table \ref{table:thermo_table}). A larger increase in RH occurs near the top of the free troposphere associated with the upward shift in the tropopause, which shifts more humid air upwards at heights previously occupied by very dry lower-stratospheric air. At low levels, MPI and CNRM RH decreases by 5\% and 1.5\% for the 0--1 km mean. In the transition layer, the 1--2 km RH lapse rate decreases slightly in both models (-3.9\% in MPI and -2.2\% in CNRM). Overall, the future climate responses of both models are again quite similar despite significant differences in their historical biases.

\subsubsection{Moist static energy deficit}

\begin{figure*}[!ht]
\centerline{\includegraphics[width=0.95\textwidth]{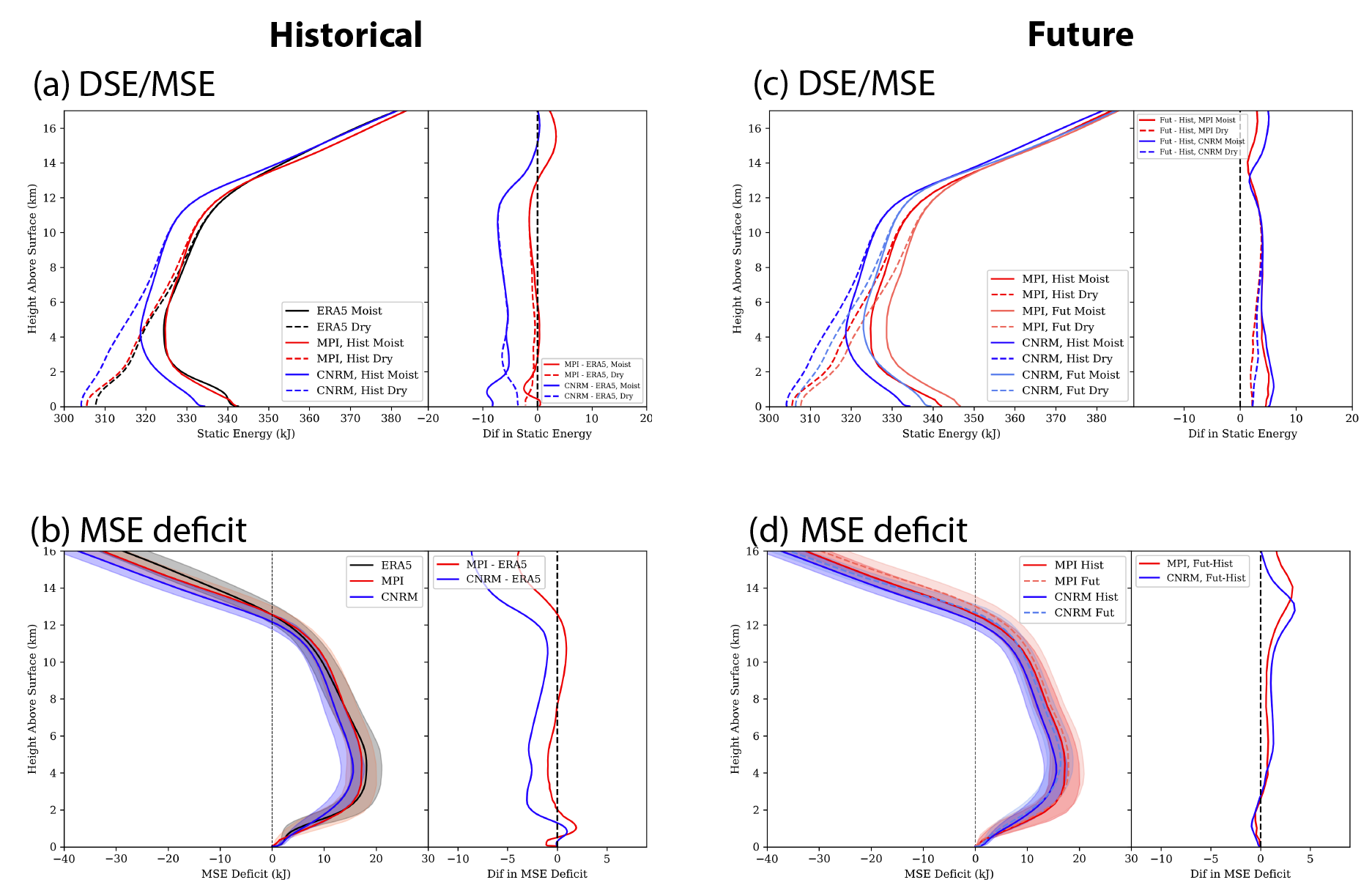}}

\caption{(a) Mean vertical profiles of dry static energy (DSE) and moist static energy (MSE) from ERA5 and the historic runs of MPI and CNRM. The shaded area represents  the interquartile range (25th--75th percentile), and the difference plot shows the differences between the ERA5 and MPI, and ERA5 and CNRM. (b) As in (a) but for moist static energy deficit, calculated for a parcel lifted from the lowest model level assuming adiabatic conservation of MSE. (c)--(d) as in (a)--(b) but for the future ssp370 and historic simulations of both models. The numbers in the key indicate sample size.}
\label{fig:MSEdef_profile}
\end{figure*}

Finally, we examine the vertical structure of the moist static energy (MSE) deficit profile due to its close link to entrainment. The MSE deficit is defined as the difference between the MSE of the parcel at the lowest model level (LML) and the MSE of the environment at each level:
\begin{equation}
MSE_{def}(z) = MSE(z_{LML}) - MSE(z)
\label{eq:MSEdef}
\end{equation}
MSE is the sum of potential, sensible, and latent energies, i.e.,
\begin{equation}
MSE = gz_p + C_pT + L_vq_v
\label{eq:MSE}
\end{equation}
where $C_p$ is the specific heat capacity of air, $L_v$ is the specific latent heat of vaporization of water, $z_p=z+z_{p,sfc}$ is the geopotential height with $z_{p,sfc}$ the geopotential height of the surface, and $q_v$ is the water vapor mass fraction (specific humidity); we neglect the latent energy content of water ice as it is generally small. Its dry counterpart, dry static energy (DSE), neglects the latent energy term. DSE and MSE are closely analogous to dry and equivalent potential temperature, respectively \citep{Betts_1974,Chavas_Dawson_2021,Chavas_Peters_2023}. The MSE deficit represents the energy difference between a parcel rising through a deep convective cloud and the surrounding near-cloud environment. To keep the calculation simple, the parcel MSE is assumed to be adiabatically conserved, thereby neglecting the MSE sink due to buoyancy \citep{Peters_Mulholland_Chavas_2022}. Entrainment is a process that mixes a parcel with the environmental air, and the energy a parcel loses per unit height is commonly parameterized in classical plume updraft models as proportional to the MSE deficit \citep[e.g.,][]{Betts_1975,Zhang_2009}. This process dilutes the buoyancy of rising air parcels and hence reduces the true CAPE realized by the parcel \citep{Peters_etal_2023}. Thus, an increase in MSE deficit would imply an increase in entrainment in deep convective clouds, all else equal.

Since the MSE deficit profile is calculated from the MSE profile itself, we begin with vertical profiles of DSE and MSE for both models and ERA5 in Figure \ref{fig:MSEdef_profile}a. Model biases in DSE profiles are unsurprisingly very similar to biases in temperature discussed previously (Figure \ref{fig:T_profile}). Biases in MSE combines the energetic effects of biases from temperature and moisture. This yields an MSE bias structure that is very similar to the DSE in the free troposphere, where latent energy content is much smaller than sensible energy, but more similar to the relative humidity bias within the boundary layer, where latent energy content is much larger. MPI reproduces the full vertical profiles of MSE and DSE remarkably well, while CNRM also reproduces their structures but with a systematic low bias that is relatively constant with height.

Mean MSE deficit profiles from both models for the historical climate are compared with ERA5 in Figure \ref{fig:MSEdef_profile}b. Both models perform well in reproducing the MSE deficit at all levels as compared to ERA5, with errors of less than 10\% of the ERA5 MSE deficit value throughout the troposphere. The largest errors are located near the boundary layer top and are likely associated with discrepancies in boundary layer height. This is particularly true for MPI due to its bias in boundary layer height (Table \ref{table:thermo_table}), whereas this bias is small in CNRM. Notably, the variability in the MSE deficit profiles is much smaller than for relative humidity (Figure \ref{fig:RH_profile}), suggesting there may be strong co-variability between boundary layer and free tropospheric MSE; this question is intriguing but lies beyond the scope of this work.

Future changes in the DSE and MSE profiles are shown in \ref{fig:MSEdef_profile}c. DSE increases relatively uniformly with height in line with changes in temperature found above. MSE also increases relatively uniformly with height, though with slightly larger increases in the lowest 2 km. This outcome is not obvious from our previous analyses, but given that MSE is approximately conserved in deep convection this suggests that to first order convection (as parameterized in the models) acts to mix MSE vertically and hence homogenize its changes in a convectively active region.

Future changes in the MSE deficit profile are shown in Figure \ref{fig:MSEdef_profile}d. Both models project relatively small changes in MSE deficit with very similar structures at all levels between models. The lone exception is the middle and upper troposphere, where CNRM projects increases of up to 10\% above 6 km while MPI changes are closer to zero. Below 4 km, both models project nearly identical changes, with a local maximum decrease in MSE deficit at approximately 1.5 km. The largest magnitude changes occur near the tropopause associated with the upward shift in the depth of the free troposphere. The modest increase in MSE deficit in the free troposphere is consistent with the finding that the MSE profile increases relatively uniformly with height as noted above, with only a slightly larger increase in MSE within the boundary layer.

Overall, across all thermodynamic variables the projected changes are quite robust between our two model simulations despite differences in their representation of the historical climate. This outcome provides greater confidence in the robustness of the projected changes.

\subsection{Changes in kinematic structure}

\begin{figure*}[!ht]
\centerline{\includegraphics[width=0.8\textwidth]{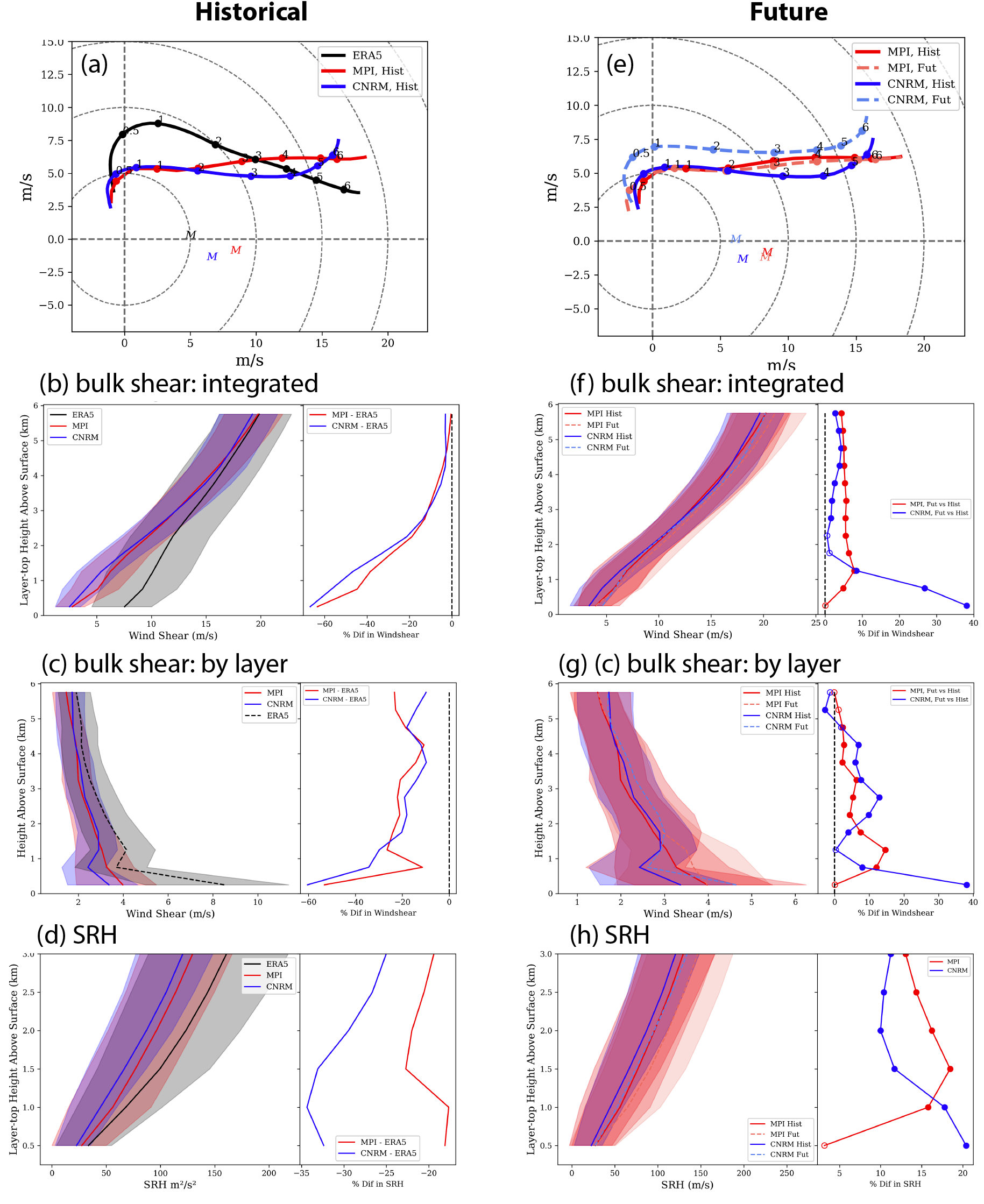}}
    
\caption{(a) Mean hodographs for ERA5 and the historic runs of MPI and CNRM. Dots correspond in order to the altitudes (0.5, 1, 2, 3, 4, 5, and 6 km), and the Bunkers storm motion is shown plotted as an 'M'. (b) Mean bulk shear integrated from the surface up to layer-top altitudes up to 6 km, for ERA5 and the historic runs of MPI and CNRM. (c) As in (b) but for the vertical profile of bulk shear calculated layerwise every 500 m (plotted point is at layer center). (d) As in (b) but for SRH up to 3 km. (e)--(h) As in (a)--(d) but for the future ssp370 and the historic runs of MPI and CNRM. The shaded area represents the interquartile range (25th--75th percentile), and the difference plot shows the differences between the ssp370 future and historic simulations of both models. For the differences in (f)-(h), a filled circle indicates the means are statistically significantly different at the 95 percent confidence level.}
\label{fig:kinematic}
\end{figure*}

We now examine the vertical structure of the lower tropospheric winds associated with SCS environments in Figure \ref{fig:kinematic}. For all of the analyses below, values of some key quantities, including those relevant to the framework of \cite{Chavas_Dawson_2021}, are provided in Table \ref{table:wind_table} for additional reference.

Mean hodographs from both models for the historical climate are compared with ERA5 in Figure \ref{fig:kinematic}a. MPI and CNRM are both quite similar to one another, and both reproduce the qualitative L-shaped vertical structure of the hodograph common in SCS environments \citep{Guarriello_Nowotarski_Epifanio_2018,Coffer_Taszarek_Parker_2020,Chavas_Dawson_2021} and also found in ERA5. This structure is characterized by a boundary layer of predominantly southerly shear in the lowest 0.5 km, with the southerly component of the flow increasing in magnitude with height. Overlying this boundary layer flow is a layer of predominantly westerly shear as the flow transitions from principally southerly to southwesterly moving upwards to 6 km altitude. In ERA5, the shear in the lower free-troposphere has a northerly component that results in a sharper wind direction shift moving across the boundary layer top before becoming unidirectional similar to the models. Both simulations have a weaker southerly wind speed in the lowest 2 km, resulting in a hodograph that is shifted slightly to the south relative to ERA5. However, the Bunkers storm motion vector of both models are similar to one another and are also shifted southeastward of the ERA5 vector by approximately 2 ms$^{-1}$. Hence, in a storm-relative sense the biases in hodograph and storm motion partially offset one another in the lowest 3 km. The surface flow vector is also nearly identical in both models. The lone notable difference between the two model hodographs is the shear between 0.5 km and 2 km, where the wind vector changes more rapidly within the 0.5--1 km layer in MPI and within the 1--2 km layer in CNRM. This difference alters the shear distribution within the 0.5--2 km layer, which we discuss next.

Figure \ref{fig:kinematic}b shows bulk shear from the surface up to layer-top altitudes from 500 m to 6 km altitude, calculated from the hodographs in Figure \ref{fig:kinematic}a. Figure \ref{fig:kinematic}c displays the vertical profile of bulk shear (units $m \: s^{-1}$) calculated layerwise within 500-m depth layers. The former directly visualizes the integrated bulk shear over any desired layer depth, while the latter visualizes the vertical distribution of shear biases (note though that integrated bulk shear biases need not equal the sum of the layerwise biases since shear is a vector difference). In each case, we show differences relative to ERA5 as percentages to compare the magnitude of the change across different layer depths. In ERA5, integrated bulk shear is large near the surface ($8 \; m \: s^{-1}$ at 500 m) and then increases at a relatively constant rate with height, from $10 \; m \: s^{-1}$ at 1 km up to nearly $20 \; m \: s^{-1}$ at 6 km. This behavior is also evident in the layerwise bulk shear profile that decreases rapidly from $8 \; m \: s^{-1}$ in the lowest 500 m to $4 \; m \: s^{-1}$ between 1 km and 2 km and then further decreases gradually towards $2 \; m \: s^{-1}$ above 4 km. Both models are similar to one another in reproducing this overall structure, consistent with their very similar hodographs, but both substantially underestimate low-level shear (50\% underestimation for 0--500 m shear). CNRM underestimations are slightly larger in the lowest 1.5 km relative to MPI owing to the different distribution of shear within 0.5--1.5 km described above.  The models also consistently underestimate layerwise shear by 20\% above 2 km, though these biases translate to relatively small biases ($<5\%$) when integrated in deeper-layer shear of 5-6 km (Figure \ref{fig:kinematic}b). Hence, both models do very well in reproducing S06 yet exhibit large low biases in bulk shear for layers closer to the surface.

Similarly, Figure \ref{fig:kinematic}d shows SRH up to layer-top altitudes from 0.5--3 km, calculated from the hodographs in Figure \ref{fig:kinematic}a. In ERA5, SRH gradually increases with increasing layer-top altitude as expected given the two-layer structure of the hodograph. The results are again qualitatively similar for the two models. Both models have a moderate low bias in SRH relative to ERA5 at all levels. The bias is smallest for the 0--3km layer (-20\% and -25\% for MPI and CNRM, respectively) and remains relatively constant across depths for MPI but increases in magnitude moving towards shallower layers for CNRM with a 30--35\% underestimation for layers up to 1.5 km. Since SRH is equal to twice the vector area of the hodograph layer relative to the storm motion vector, the low bias in SRH in the two models arises due to the weaker southerly component of the flow in the lowest 1 km of their hodographs relative to ERA5 in Figure \ref{fig:kinematic}a. This difference can also be interpreted as due to the low bias in shear in the lowest 1 km seen in Figure \ref{fig:kinematic}c, as this low shear bias is effectively integrated upwards with height to yield the low biases in SRH, an effect amplified for the shallowest layers.

Future changes in the simulated hodographs in each model are shown in Figure \ref{fig:kinematic}e. In both models, the hodograph shifts west/northwestward within the lowest 1 km, resulting in slightly stronger curvature. Meanwhile, the shear structure above 1 km remains nearly constant relative to the 1 km flow. In CNRM, the entire hodograph is to first-order simply translated, with a northwestward shift at lower levels and more northward shift at higher levels. In MPI, the hodograph barely shifts at all except for a westward shift near the surface. The net result is a slightly enhanced curvature in the lowest 1 km and shear in the lowest 2 km. The storm motion vector shifts northwestward in CNRM and very slightly southward in MPI, both consistent with the shifts in their respective hodographs. Hence, for storm-relative flow, the shift in the hodograph is partially offset by the shift in the storm-motion vector.

The change in integrated bulk shear is shown in Figure \ref{fig:kinematic}f and changes in the vertical distribution of bulk shear is shown in Figure \ref{fig:kinematic}g, along with the percentage differences in future relative to historical for each. Recall that for our analysis we restricted our range of S06 values to focus on future changes in vertical structure for a given value of bulk environmental parameters. Unsurprisingly, then, we find that the 0--6 km wind shear changes only minimally in both models, with small increases of approximately 5\%. The percent change is relatively small for all layer-top altitudes greater than 2 km and consistent in both models, with near zero change for layer-top altitudes of 2-3 km in CNRM. In contrast, within the lowest 2 km, integrated bulk shear increases more strongly in both models, though the models differ in magnitude and altitude of peak change: both models show +10\% at 1.5 km, but for shallower layers this value decreases towards zero in MPI whereas it increases to a peak of +40\% for the shallowest layer (500m) in CNRM. This strong increase in near-surface shear in CNRM is associated with the stronger northwestward shift in the hodograph at 0.5 km relative to the surface noted above. These outcomes are also evident in the changes in layerwise bulk shear profiles, with the lone notable difference being that CNRM exhibits more substantial increases in shear within the 2-5 km layer associated with a subtle reduction in the curvature of the hodograph (Figure \ref{fig:kinematic}a), though these layerwise directional shear changes have a relatively small impact when integrated over a deeper layer such as for S06. Overall, the wind shear changes over all layer depths are modest, though with some indication of a stronger increase in shear (corresponding to a larger percentage change as well) within the lowest 1 km. However, the magnitude and particularly vertical structure of those changes differ markedly between our two models, indicating significant uncertainty in this outcome. Moreover, it is important to note that the model biases relative to ERA5 were also largest in the lowest 1 km, which may further reduce confidence in the model outcomes despite the consistency between models.


The change in SRH is shown in Figure \ref{fig:kinematic}h, along with the percentage differences in future relative to historical. The qualitative structure of the changes in SRH mirror the changes in integrated bulk shear discussed above (Figure \ref{fig:kinematic}f). Both models show increases in SRH over all layer depths in the range of +10--20\%. Changes are smaller for the 0--3km layer (+10--15\%) in both models. For shallower layers the models differ, with MPI showing the largest increase for the 0--1.5 km layer (+20\%) while CNRM showing SRH increasing more strongly for the shallowest layers (+20\% for the 0--500m layer). These increases in SRH are associated with the increases in shear found primarily within the lowest 1.5 km but whose detailed structure differs between the two models. Such changes in the low-level shear structure are effectively integrated upwards in the calculation of SRH. Ultimately, both models show consistent moderate increases in both 0-1 km and 0-3 km SRH, though they give highly divergent behavior for 0-500 m SRH, again reflecting significant uncertainty in the representation of the detailed structure of wind shear within the lowest kilometer.


\begin{table}[]
\caption{Key quantities calculated from the mean hodographs from ERA5, MPI, and CNRM, including those relevant to the SCS sounding model of \cite{Chavas_Dawson_2021}. Values are the mean with interquartile range (25th-75th percentile) in parentheses.}
\centering
\begin{adjustbox}{max width=0.5\textwidth}
\begin{tabular}{|ll|l|l|l|l|l|l|l|}
\hline
 &  & ERA5 &  & \multicolumn{2}{l|}{MPI-ESM1-2-HR} &  & \multicolumn{2}{l|}{CNRM-ESM2-1} \\ \cline{5-6} \cline{8-9} 
 &  &  &  & HIST & FUT &  & HIST & FUT \\ \cline{1-3} \cline{5-6} \cline{8-9} 

\multicolumn{2}{|l|}{Usfc (ms$^{-1}$)} & -0.8 (-2.4, 0.6) &  & -0.8 (-2.6, 0.8) & -1.4 (-3.2, 0.3) &  & -1.1 (-3.1, 1.0) & -1.5 (-3.3, 0.3) \\ \hline
\multicolumn{2}{|l|}{Vsfc (ms$^{-1}$)} & 3.7 (1.5, 6.2) &  & 2.3 (-1.0, 5.7) & 1.8 (-1.6, 5.1) &  & 2.5 (-0.1, 5.1) & 2.8 (0.1, 5.7) \\ \hline
\multicolumn{2}{|l|}{Total Shear 0--Hbl (ms$^{-1}$)} & 9.4 (5.4, 12.6) &  & 3.4 (1.5, 4.7) & 4.9 (2.5, 6.7) &  & 4.0 (1.9, 5.5) & 5.7 (2.7, 7.7) \\ \hline
\multicolumn{2}{|l|}{Zonal 0--Hbl (ms$^{-1}$)}       & 2.2 (-1.2, 5.5) &  & 0.2 (-0.9, 1.1) & 0.9 (-0.9, 2.6) &  & 1.3 (0.1, 2.5) & 1.7 (-0.4, 3.3) \\ \hline
\multicolumn{2}{|l|}{Meridional 0--Hbl (ms$^{-1}$)}  & 5.1 (0.0, 10.3) &  & 2.1 (-0.1, 4.0) & 3.1 (0.2, 5.7) &  & 2.9 (0.7, 4.5) & 4.1 (0.9, 6.7) \\ \hline
\multicolumn{2}{|l|}{Total Shear Hbl--3km (ms$^{-1}$)} & 11.6 (8.6, 14.5) &  & 12.0 (9.2, 14.3) & 12.0 (9.0, 14.8) &  & 11.1 (8.5, 13.4) & 10.8 (8.1, 13.3) \\ \hline
\multicolumn{2}{|l|}{Zonal Hbl--3km (ms$^{-1}$)}       & 8.8 (5.5, 12.0)  &  & 10.0 (7.4, 12.4) & 9.9 (6.9, 12.6) &  & 9.2 (6.4, 11.8) & 8.5 (5.7, 11.3) \\ \hline
\multicolumn{2}{|l|}{Meridional Hbl--3km (ms$^{-1}$)}  & -2.7 (-7.5, 1.9) &  & 1.6 (-2.8, 5.9) & 0.7 (-4.2, 5.4) &  & -0.5 (-4.7, 3.6) &  -0.4 (-4.8, 4.1) \\ \hline
\multicolumn{2}{|l|}{Total Shear 3--6km (ms$^{-1}$)} & 9.3 (6.2, 12.0) &  & 9.2 (6.8, 11.2) & 9.4 (6.8, 11.7) &  & 8.7 (6.2, 10.7) & 9.0 (6.4, 11.2) \\ \hline
\multicolumn{2}{|l|}{Zonal 3--6km (ms$^{-1}$)}       & 6.4 (3.0, 9.5)  &  & 7.3 (4.8, 9.7) & 7.5 (4.8, 10.0) &  & 6.3 (3.7, 9.1) & 6.7 (4.0, 9.2) \\ \hline
\multicolumn{2}{|l|}{Meridional 3--6km (ms$^{-1}$)}  &-2.2 (-6.3, 1.7) &  & 0.1 (-3.5, 3.6) & 0.5 (-3.3, 4.0) &  & 1.6 (-2.0, 5.2) & 1.5 (-2.3, 5.2) \\ \hline

\end{tabular}
\end{adjustbox}
\label{table:wind_table}
\end{table}

Overall, for the kinematic fields the two models behave quite similarly to one another in both their representation of the historical climate (and hence their biases too) and in their projected changes in these fields with future climate change. Both models project significant changes in the hodograph principally in the lowest 2 km. These changes translate to projected increases in shear below 2 km, though with minimal changes in 0--6 km shear. The changes in shear translate to projected increases in SRH at all levels, with similar moderate increases for 0-3 km and 0-1 km levels but only one model shows a comparable increase for 0-500 m.

\subsection{Implications}

Taken together, two notable changes emerge: potential increases in low-level shear and in MSE deficit. Otherwise, the free troposphere becomes deeper while maintaining a relatively constant tropopause temperature and lapse rate, indicating a thermodynamic environment that remains highly conducive to buoyant deep convective updrafts.

Given recent work finding that tornadogenesis depends most strongly on shear in the lowest 1 km and possibly even lowest 0.5 km \citep{Coffer_etal_2019,Coffer_Taszarek_Parker_2020}, the increase in low-level shear suggests that the SRH ingredient for tornadogenesis could be enhanced when supercells form from a fixed bulk environment. However, other less well-understood factors that influence tornadoes (e.g., updraft width) may change with warming as well that could potentially offset the changes in shear structure analyzed here. Moreover, the poor representation of low-level shear in the historical climate as compared to ERA5, coupled with the lack of clear agreement in the structure of its projected change between our two models, indicate that this conclusion must be made with substantial caveat.

Meanwhile, an increase in MSE deficit suggests the possibility of enhanced entrainment dilution. An increase in entrainmment dilution may result in a reduction in the frequency of severe convective storms by reducing the true parcel CAPE from its undiluted value \citep{Peters_etal_2020a,Peters_etal_2023}. The above interpretation is consistent with recent high-resolution regional modeling studies that find an increase in the frequency of SCS environments yet a decrease in the frequency of SCS activity over the central Great Plains region studied here \citep[][]{Ashley_Haberlie_Gensini_2023}. This finding is typically ascribed to increases in convective inhibition, though changes in entrainment may play a similar role and is worthy of deeper study. Additionally, the slight decrease in boundary layer relative humidity may increase LCLs, which may exacerbate these entrainment effects.

Overall, evaluation of the net response of severe thunderstorms and tornadoes themselves is too complex to be ascertained here. Nonetheless, we hope this work provides a basis from climate model projections to build upon in future work. 


\section{Conclusions}\label{sec:conclusions}

Recent evidence suggests that severe thunderstorms and possibly tornadoes may become more frequent and/or intense in the future. Our understanding of this behavior is typically explained via changes in common vertically integrated (i.e., ``bulk'') variables, such as CAPE and 0--6 km shear. However, the vertical structure of thermodynamic and kinematic profiles in severe convective storm environments possess many more degrees of freedom that can change independently of these standard bulk parameters. This work examined how climate change may affect the complete vertical structure of these environments for a fixed range of values of CAPE and S06, using soundings over the central Great Plains from two high-performing climate models for the high-end forcing ssp370 scenario. Hence, our results may be thought of as probing future climate changes in severe convective storm environments \textit{not} associated with changes in CAPE and S06.

We summarize our primary results as follows:
\begin{enumerate}

\item Changes in thermal structure: temperature profiles warm relatively uniformly with height, free tropospheric lapse rates decrease slightly, and the free tropopause shifts upwards at approximately constant temperature.

\item Changes in moisture structure: For relative humidity, the boundary layer becomes slightly drier (-2--5\%) while the free troposphere becomes slightly moister (+3--5\%). The boundary layer height remains constant.

\item Changes in moist static energy deficit: free-tropospheric moist static energy deficit increases (1--10\%) above 4 km altitude, suggesting that the effects of entrainment could become stronger with warming, though the models do not agree on the magnitude. This outcome is consistent with the finding that MSE itself increases relatively uniformly with height with a slightly larger increase within the boundary layer.

\item Changes in kinematic structure: Hodographs become more strongly curved, due to stronger southerly/southeasterly flow between 500 m and 1.5 km relative to flow near the surface and above 3 km that remains largely unchanged. This behavior results in stronger wind shear within the lowest 1.5 km and greater storm-relative helicity within the lowest 1.5 km (which enhances SRH through all layers up to 3 km).

\item Changes in both thermodynamic and kinematic profiles are relatively consistent between our two models, despite different bias structures relative to ERA5, suggesting that the qualitative outcomes are robust.

\item Overall, the most notable change is the increase in low-level shear and SRH that may suggest increased potential for severe thunderstorms and tornadoes forming within high CAPE and high shear environments in our domain. However, modest changes in other factors, including both the slight increase in free tropospheric MSE deficit, which may enhance entrainment effects, and the slight decrease in boundary layer relative humidity, which may increase LCLs, may offset these effects. Evaluation of the net response of severe thunderstorms and tornadoes themselves is too complex to be ascertained here.

\end{enumerate}

The findings of a slight reduction in free tropospheric lapse rate and boundary layer relative humidity with warming within our fixed ranges of CAPE and S06 is consistent with similar findings in \cite{Wang_Moyer_2023} for the entire distribution over summertime North America. We reiterate that such changes would be on top of changes associated with the expected large increases in CAPE with warming that would likely increase the intensity, and possibly frequency, of severe convective storms as noted in past studies.

Our interpretation is by no means final, but rather should be tested in numerical simulations and placed in the context of other climate-driven changes in severe convective storm environments. Our effort is a starting point for considering the full vertical structure of the thermodynamic and kinematic environment from climate models to better understand how severe thunderstorms and tornadoes may change with climate change. Ideally, this information would be integrated into a climate-dependent theory for these phenomena. Future work should test outcomes in limited-area numerical model experiments of individual storms. Moreover, our analysis can be extended to other geographic regions, particularly the southeast United States, where the nature of severe thunderstorm environments and events are known to differ in complex ways from the Great Plains \citep{Sherburn_etal_2016}. This carries added importance given the recent shift in tornado activity towards the southeast U.S. \citep{Gensini_Brooks_2018}. The outcomes can be compared against regional model experiments that can properly represent convective initiation and the array of convective forcing agents found in the real world. Additionally, future work may want to consider the components of storm-relative helicity, particularly environmental horizontal vorticity and storm-relative flow, given emerging research that it is specifically these two components of SRH that are associated with intense low-level mesocyclogenesis \citep{Goldacker_Parker_2023}.

Finally, we note that here we do not offer explanations for \textit{why} key aspects of the vertical structure, such as the curvature of the low-level hodograph or the free-tropospheric relative humidity, do or do not change in the future. This type of understanding requires a broader analysis of how the large-scale circulation pattern will change and how this interacts with the land surface over the continental interior. Such endeavors are highly complex but also worthwhile, especially for understanding climate change impacts on tornadoes given the critical importance of the near-surface wind structure for tornadogenesis.

\acknowledgments
The authors thank three anonymous reviewers for thorough and highly constructive feedback that greatly improved this manuscript. The authors were supported by National Science Foundation (NSF) AGS grants 1648681 and 2209052 and NASA grant 19-EARTH20-0216. We also acknowledge the open-source Python community, and particularly the authors and contributors to the Matplotlib \citep{matplotlib}, NumPy \citep{numpy}, and MetPy \citep{metpy} packages that were used to generate many of the analyses and figures. Other post-processed data are available from the authors upon request.

%
%
\datastatement


6-hourly  ERA5 reanalysis data were accessed on model levels from \url{https://doi.org/10.5065/XV5R-5344} and for the near-surface and on pressure levels from \url{https://doi.org/10.5065/BH6N-5N20} \citep{NCAR_RDA_ERA5}. 6-hourly CMIP6 model historical and future experiment (ssp370,ssp585) data were accessed from \url{https://esgf-node.llnl.gov/search/cmip6}. Analyses were performed on the NCAR Cheyenne and Casper supercomputers \citep{NCAR_Cheyenne} as well as on computational resources provided by Purdue Rosen Center for Advanced Computing \citep[RCAC; ][]{Purdue_RCAC}.

\bibliographystyle{ametsocV6}
\bibliography{refs_SLSEARTH,refs_CHAVAS}

\end{document}